\documentstyle[preprint,prb,aps]{revtex}
\begin{document}
\draft
\title{Resonating-Valence-Bond Ground-state of CaV$_4$O$_9$ \\
Studied by the Gutzwiller-projected Schwinger-boson method}
\author{T.~Miyazaki and D.~Yoshioka}
%%%%%%%%%%%%%%%%%%%%%%%%%%%%%%%%%%%%%%%%%%%%%%%%%%%%%%%%%%%%%%%%%%%%%%%%%%%%%
\address{Department of Basic Science, University of Tokyo\\
3-8-1 Komaba, Tokyo 153, Japan\\}
%%%%%%%%%%%%%%%%%%%%%%%%%%%%%%%%%%%%%%%%%%%%%%%%%%%%%%%%%%%%%%%%%%%%%%%%%%%%%
\date{\today}
\maketitle
\begin{abstract}
An antiferromagnetic Heisenberg model on a
 1/5-depleted two-dimensional square-lattice, a model of CaV$_4$O$_9$,
is investigated by variational Monte Carlo simulation.
A prototype of a trial wave function is made by
projecting out the doubly occupied states
from the Schwinger-boson mean-field solution.
Then variational Monte Carlo simulation is performed
up to $40 \times 40 $ sites(including $320$ vacant sites).
The optimized state has the lowest energy, $-0.5510J$, ever reported.
For this state energies of a dimer bond and a plaquette bond,
staggered magnetization, static structure factor, and
excitation spectrum are calculated.
It is shown that the N\'eel order survives and there is no gap
at isotropic coupling.
\end{abstract}
\pacs{75.10.-b, 75.10.Jm, 75.50.Ee}
%
%
%
%------------ 1. INTRODUCTION----------
Recent discovery of the energy gap in CaV$_4$O$_9$ has given rise to the
question of the origin of the two-dimensional spin-gap.\cite{Tani}
Though it has been suggested that the pseudo-gap is observed
for the underdoped bilayer YBCO system,\cite{NMR,newtron}
this is the first material to show the gapful
ground-state for spin system in the
two-dimensional single layer experimentally.
Considering the fact that the two-dimensional
square-lattice Heisenberg layer has an
antiferromagnetic long-range order,\cite{Ma}
the periodic lattice-defects intrinsic to the material
 may play an important role to destroy the magnetic
ordering to make the spin-gap.\par
%
%------------1.1.2 CaV4O9 Lattice structure and relevant electron-------
The planer lattice-structure of CaV$_4$O$_9$ is shown in Fig.1.
Each closed circle stands for the V atom which has a localized spin
with $S=1/2$.
It interacts with the nearest neighbor spins antiferromagnetically.
Though topologically there are two
different types of bonds, dimer bond $J_0$ and plaquette bond $J_1$,
both couplings are believed to have the same magnitude in the material.
Thus in this rapid communication,
we consider only the
model with $J \equiv J_0=J_1$.
In addition to the above coupling, it has also been pointed out that
in the real material there is a frustrated
next nearest neighbor coupling of the order of $J'=J/2$,\cite{Ueda2}
which we neglect in the present work.
Thus we investigate here the possibility to have a ground-state with
spin-gap in a system represented by the following Hamiltonian,
\begin{equation}
H = J  \sum_{i} \sum_{w} \mbox{\boldmath $S_{i} \cdot S_{i+w}$}.
\end{equation}
Here, ${\boldmath i}=(i_x,i_y)$, $i_x$ and $i_y$ being integers,
${\boldmath i+w}=(i_x+1,i_y)$ or $(i_x,i_y+1)$,
and the summation over
${\boldmath w}$ is done only for sites where spin exists.
\par

%------------1.2  The previous investigations ----------
There has been several investigations to explain the experimental
spin-gap.\cite{Ueda2,Katoh,Ueda1,SBMFT,Sano,Ueda3}
The essential problem of this CaV$_4$O$_9$ is
whether the gap is made by
the depletion of the square-lattice or the additional frustration.
Imada and Katoh\cite{Katoh}
have proposed that the gap is explained by the
resonating local singlets of the four spins on a plaquette,
which they call plaquette RVB state.
They investigated this model by Quantum Monte Carlo (QMC) method and
the second order perturbation expansions (PE)
from the plaquette limit. It is found
that the system has a gapful
excitation without frustration.
The estimated spin-gap is
$\Delta=0.11 \pm 0.03 J$ for QMC and $0.205J$ for PE.
On the other hand, linear spin-wave theory and Schwinger-boson mean-field
approximation resulted in the gapless excitation
as far as frustrated couplings are not included.\cite{Ueda1,SBMFT}
In the latter method it has been found that 
the N\'eel state is stable at $ 0.60  < J_0/J_1 < 2.40$.
Contrary to ref.(\onlinecite{Katoh}) these results
suggest that the next nearest neighbor coupling is needed to make the
gap.
Exact diagonalization study gives a contradictory
result.\cite{Sano}
It shows a small spin-gap with a finite staggered
magnetization. This may result from the difficult
extrapolation to the infinite size.
Quite recently, QMC simulation done by
Troyer {\it et al}.\cite{Ueda2} shows there is no gap without frustration.
They make a phase diagram as a function of the ratio of
dimer coupling $J_0$ to plaquette one $J_1$.
In both limits of $J_0 \gg J_1$ and $J_1 \gg J_0$,
the system has a gapful state.
Thus ordered state is realized only in a limited region of the
phase diagram.
The estimated ordered region is at
$ 0.60 \pm 0.05 < J_0/J_1 < 1.07 \pm 0.05$.
This shows that the isotropic point is quite close
to the order-disorder transition and suggests that other kinds
of investigations are needed for this isotropic coupling.
\par

In this paper, we study this model by Gutzwiller-projected
Schwinger-boson method.\cite{Chen,MYO,MNY}
In this method, first, the Hamiltonian
is solved by Schwinger-boson mean-field theory.\cite{Arovas,aa,yd1,yd2}
Then the solution is Gutzwiller projected to fix the spin at each site
to be $S=1/2$.
The obtained ground-state is a kind of long-range RVB \cite{RVB1,RVB2,LDA}
state with amplitude for a constituent bond depending on the distance
between the sites.
The mean-field solution is used only for the determination
of these bond amplitudes.
As for the mean field solution we could use that by
Albrecht and Mila.\cite{SBMFT}
However, for simplicity we here use the bond amplitudes obtained from
the solution of the simple square-lattice,
except those for the nearest neighbor bond amplitudes.
Namely, in this paper, we modify the plaquette and dimer bond
amplitudes by scaling them by $\alpha_{\rm p}$
and $\alpha_{\rm d}$, respectively,
where $\alpha_{\rm p}$ and $\alpha_{\rm d}$ are variational parameters.
%since the weight of the short-range RVB strongly contributes to the
%wave function.

This type of Gutzwiller-projected wave functions have already given
some good results for other models.
It was first applied to the square-lattice Heisenberg model.\cite{Chen}
The resultant ground-state is quite close to the true ground-state:
the energy agrees with the best estimated value\cite{Series,GFMC}
within the error of less than 0.1\%.
This method has then been applied to other similar models
and also gave good results for each of them.\cite{MYO,MNY}
We will see later in this paper that the ground state energy
is low enough in this system also.
The optimized wavefunction is used to calculate
the staggered magnetization and the excitation spectrum.
\par

%Especially, in our recent paper, we found the way to calculate
%the excitation spectrum by single mode approximation.\cite{MNY}
%A merit of this method is to
%calculate the dispersion relation in an entire Brillouin zone.
%We can calculate the magnitude of the gap for a gapful system
%or the gapless point for a gapless one.
%\par
%
%
%
%
%
%
%-------------------{4.Numerical results}-------------------------------
Using these methods, we did the variational Monte Carlo simulation
for systems with $10 \times 10 $, $20 \times 20 $,
$30 \times 30 $, and $40 \times 40 $ lattice points.\cite{vacancy}
The variational parameters are the ratios $\alpha_{\rm p}$
and $\alpha_{\rm d}$.
It was found that
the optimal state is realized at $\alpha_p=1.08$ and  $\alpha_d=0.93$.
This anisotropy is strongly enhanced
in the result of the energy
for the dimer coupling per bond($E_d$),
and that for the plaquette one($E_p$).
Actually, they
take quite different values: $E_d=-0.303J$, $E_p=-0.400J$.
This shows that the system favors the plaquette RVB state
even in the isotropic coupling.
The total energy per site can be scaled by $1/L^3$
as shown in Fig.2.
Here, L is the lattice size.
It is clear from Fig.2 that
our system size is large enough to estimate the energy
in the thermodynamic limit.
The obtained optimal energy is
$-0.5510 \pm 0.0005$ per site, which is the lowest value
that has already been reported.
% and guarantee our trial wave function.
\par
%------------4-2. How to determine the gapless system----------
To determine whether the system has gap or not without frustrated
coupling, we have calculated staggered magnetization
and excitation spectrum.
The result of staggered magnetization per site can be scaled by
inverse lattice-size $1/L$.\cite{Huse}
The magnetization is evaluated from
the longest spin-spin correlation as,
\begin{equation}
   \label{eqn:M}
M(L)   \equiv   \sqrt{\frac{1}{N} {\sum_{i,j}}^{'}
     \langle | \mbox{\boldmath $S$}_{i} \cdot \mbox{\boldmath $S$}_{j}  |
                 \rangle}  \quad,
\end{equation}
where the summation is taken over the longest possible distance
in the plane and $N$ is the number of spin sites.
>From this analysis we have concluded that
there is an antiferromagnetic long-range order.
The staggered magnetization per site in the thermodynamic limit is
$0.25 \pm 0.01$.
\par
Although the existence of the long-range order indicates the absence of
excitation gap, we also
confirmed this by calculation of the excitation spectrum.
We use the single mode approximation.\cite{MNY,Fyn1,Fyn2}
Since this is a variational calculation, the upper bound of the spectrum
is obtained.
The true unit cell of this system is defined by the lattice vectors,
${\boldmath a_1}=(2,1)$ and ${\boldmath a_2}=(-1,2)$, and contains
four spins.\cite{2112}
Therefore in the Brillouin zone there are four spin excitation modes.
{}From the experience in the simple square-lattice, we expect that
the lowest modes among the four are those
where the nearest neighbor spin moves almost in-phase or out-of-phase.
Such modes can be expressed by
\begin{equation}
|\mbox{\boldmath $k$} \rangle = \sum_i S_{i}^-
e^{i\mbox{\boldmath $k$} \cdot \mbox{\boldmath $r_{i}$} }|G\rangle,
\label{eqn:exci}
\end{equation}
where $|G\rangle$ is the ground state, and the value of the wave vectors
are around $\mbox{\boldmath $k$} = (0,0)$ or
around $\mbox{\boldmath $k$} = (\pi,\pi)$, i.e.
the wave vectors around the $\Gamma$ or $M$ points of the
original square-lattice.
The upper bound of the spectrum,
$\omega(\mbox{\boldmath $k$})$,
is expressed as
\begin{equation}
\label{eqn:Ek}
\omega(\mbox{\boldmath $k$}) =
\frac{f(\mbox{\boldmath $k$})}
{S(\mbox{\boldmath $k$})} \quad ,
\end{equation}
\begin{equation}
\label{eq:Sk}
S(\mbox{\boldmath $k$})  =
\frac{1}{N} \sum_{i,j}
\langle G | S_{i}^+ S_{j}^- | G \rangle
e^{i\mbox{\boldmath $k$} \cdot \mbox{\boldmath $r_{i,j}$} } \quad ,
\end{equation}
\begin{eqnarray}
\label{eq:fk}
f(\mbox{\boldmath $k$}) & = & \frac{J}{N} (2-\cos{k_x}-\cos{k_y})
\sum_{i,w'}
\langle G | S_{i+w'}^+ S_{i}^- S_{i+w'}^z | G \rangle .  \nonumber \\
& &
\end{eqnarray}
Here,
$i+\omega^{'}=(i_x \pm 1, i_y)$ or
$(i_x, i_y \pm 1)$,
$S(\mbox{\boldmath $k$})$ is the static structure factor,
and $f(\mbox{\boldmath $k$})$ is a 3-point correlation
function of spin operators, which is proportional to the
ground-state energy.\cite{MNY}
Static structure factor, $S(\mbox{\boldmath $k$})$,
and the excitation spectrum, $\omega(\mbox{\boldmath $k$})$, for
$40 \times 40$ lattices
are shown in Figs.3 and 4, respectively.
Here, $\Gamma=(0,0)$, $X=(\pi,0)$, and $M=(\pi,\pi)$ in the momentum
space.\cite{Note3}
Error bar is the statistical error.
The value of $S(\mbox{\boldmath $k$})$ at $M$ point is too large
to be plotted in this figure.
The structure factor also suggests the strong
antiferromagnetic correlation.
The excitation is gapless at the $\Gamma$ and $M$ points.
It should be noted that in the true Brillouin zone this $M$ point
is folded into the $M$ point of the true Brillouin zone,
which is at $(\pi/5,3\pi/5)$.
%Namely we take the following state as a trial state for the lowest
%energy state with momentum $\mbox{\boldmath $k$}$:
%Thus we consider the same spin wave as the original square-lattice,
%where vacancies are filled by spins, and the allowed values of
%$\mbox{\boldmath $k$}$ are those within the same Brillouin zone
%as the original lattice.
%Of course these wave vectors are not independent with each other
%in the present system.
%The periodic vacancies reduce the Brillouin zone 1/10 of the original
%one in the area.
%The spectrum obtained should be folded into the true Brillouin zone,
%and coupling between the equivalent wavevectors will shift the spectrum,
%and reduce the number of modes to eight.
%However, we dare show the results in the large Brillouin zone
%to compare the results with that for the square-lattice.
%We expect that in the low energy, long wave length limit
%the coulping with other wave vectors are small,
%and our approximation gives reasonable results.
%This is because there the energy difference makes the coupling small,
%and the effect of the vacancies are averaged out.
\par
%-------------------{4.Discussion}-------------------------------%
%--------4-1.energy-----justify out VMC-------------
To discuss the detailed properties of our results,
we first see the accuracy of our trial wave function.
The energy per site of the obtained ground-state is quite good.
As for the energy, there is no elaborated estimation.
It is easily estimated that the energy per site
for an isolated plaquette RVB state is $-0.5J$.\cite{Ueda1}
In the presence of inter-plaquette coupling,
the energy should be lowered.
By the second order perturbation expansions from the plaquette RVB state,
Katoh and Imada gives it as $-0.5373J$.\cite{Katoh}
This value is quite close to the result of the
linear spin-wave theory, $-0.5376J$.\cite{Ueda1}
On the other hand,
our Monte Carlo result shows that the energy is $-0.5510 \pm 0.0005 J$
which is the lowest energy that has already been reported.
This means that the Gutzwiller-projected Schwinger-boson RVB
wave function is useful to investigate other physical properties
as well as the single-layer two-dimensional Heisenberg model.\cite{Chen}
\par
%--------4-2.Argue about the gapless result--------
Our results of the finite staggered magnetization and excitation spectrum
show gapless spin excitation at the isotropic coupling.
The result of finite magnetization coincides with
the Schwinger-boson mean-field theory.
The Gutzwiller-projection slightly changes the magnitude of the
magnetic ordering.
The magnitude of the staggered magnetization, $0.25 \pm 0.01$ per site,
is slightly larger than that of
the Schwinger-boson mean-field theory, about $0.22$.\cite{SBMFT}
The gapless points exist at the $\Gamma$ and $M$ point.
The gap closing at the former point is made by
the behavior of $S(\mbox{\boldmath $k$})$ near the $\Gamma$ point.
Since $f(\mbox{\boldmath $k$})$ is proportional to $k^2$,
linear $k$ dependence of $S(k)$, Fig.3, makes the spectrum gapless.
The antiferromagnetic long-range order makes the excitation spectrum
gapless at the $M$ point,
since structure factor diverges here.
These results show that other ingredients are necessary to
explain the experimental results on CaV$_4$O$_9$.
The most probable mechanism is
the destruction of the ordered state by additional next nearest
neighbor couplings.
Another candidate is subtle deformation of the
lattice-structure which makes the dimer or plaquette couplings
change to make the spin-gap.\cite{Ueda3}
\par
%--------4-3. The different energy between plaquette and dimer----------
Compared with the simple square-lattice, the bond energy of the plaquette,
$E_p$, is enhanced, while that of the dimer bond, $E_d$, is suppressed.
Namely $E_p$ is larger than that of the square-lattice, $-0.334J$,
and about
80\% of the pure plaquette state.
On the other hand, $E_d=-0.303J$ is reduced from $-0.334J$.
This suggests that the system is approaching the plaquette RVB state
in spite of the isotropic coupling.
This result is consistent with the phase diagram
given by QMC method.\cite{Ueda2}
There,
the isotropic point is quite close to the order-disorder transition
beyond which the gap originated by the plaquette RVB opens.
\par
%
%--------4-4.Depletion effect-----comparison with 2D square-lattice----
In the present system the energy per site, $-0.5510J$ is smaller
than that of the square-lattice, $-0.6696J$.
However, if energy is compared per bond, the average energy per bond
$-0.3674J$ of this system is lower than $-0.3348J$ of the latter.
This enhancement of the bond strength comes from the smaller coordination
number, and also indicates closeness of the present system towards the
spin-gap state.
This tendency is also seen
in the long-range spin-correlation.
The magnetization per site is reduced to $83\%$
of the square-lattice.\cite{Ma}
These results suggest that more vacancies make the system gapful.
\par
The spin wave velocity of the CaV$_4$O$_9$
is lower than that of the square-lattice.
%the discontinuity of the slope
%at $\mbox{\boldmath $k$}=(\pi/2,0)$ . This discontinuity
%is not seen in the two-dimensional Heisenberg layer.
%The depletion must make this discontinuity. However, there is
%no reason why this discontinuity exists at such point.
The renormalized factor, $Z_c$, which is rescaled by the classical
spin-wave velocity,
does not show the size-dependence.
The estimated $Z_c$ is $1.94 \pm 0.02$, which is smaller
than our result of the square-lattice, 1.99.
\cite{Note2}
The decrease of the spin-velocity can be understood
by the decrease of the coordination number,
which effectively weakens the spin-stiffness.
\par
In conclusion, we have investigated
the antiferromagnetic Heisenberg model on
a $1/5$-depleted two-dimensional square-lattice as a model of
CaV$_4$O$_9$ by variational Monte Carlo simulation.
In our model, where the interaction exists only between the
nearest neighbor spin pairs, the antiferromagnetic
long-range order survives and the excitation is gapless.
The gapless points exist at the $\Gamma$ and $M$ points.
Plaquette energy is larger than dimer energy,
which shows the system favors the plaquette RVB state.
Our conclusion is consistent with
the recent QMC result done by Troyer {\it et al}.\cite{Ueda2}
These results show that
the additional mechanism is
needed to destroy the
magnetic ordering and to make the experimental spin-gap state
of CaV$_4$O$_9$.
%
%
%---------------------------acknowledgements------------------------------

The authors thank M.~Ogata and one of the authors(T.M.) thanks N.~Katoh
and M.~Troyer for useful discussions.

%

%----------------------------Figure Caption-----------------------------
\newpage
\section*{FIGURE CAPTIONS}
FIG.~1. \quad
Lattice structure for V-spins in a layer of CaV$_4$O$_9$.
Dimer(plaquette) coupling is shown by a thick(thin) solid line.
\par
\vspace{1cm}
FIG.~2. \quad
Energy per site as a function of $1/L^3$.
The optimal parameters are taken as
$\alpha_p=1.08$ and $\alpha_d=0.93$.
The obtained value in the thermodynamic limit is
$-0.5510 \pm 0.005$.
\par
\vspace{1cm}
FIG.~3. \quad
Structure factor, $S(\mbox{\boldmath $k$})$,
for $40 \times 40$ lattice.
The value at $M$ point is too large to show in the figure.
\par
\vspace{1cm}
FIG.~4. \quad
Excitation spectrum, $\omega(\mbox{\boldmath $k$})$,
for $40 \times 40$ lattice.
The spectrum is gapless at $\Gamma$ and $M$ point.
\par
\vspace{1cm}

\begin{thebibliography}{99}
\bibitem{vacancy} The system size refers to the simple square-lattice,
thus the vacant sites are counted, i.e. system of the size $40\times40$
contains 320 vacant sites.
\bibitem{2112} If we consider the position of apical oxygen, i.e.
three dimensional structure,
the unit cell should be doubled.
Here, we only consider the planer structure of CaV$_4$O$_9$.
\end{thebibliography}
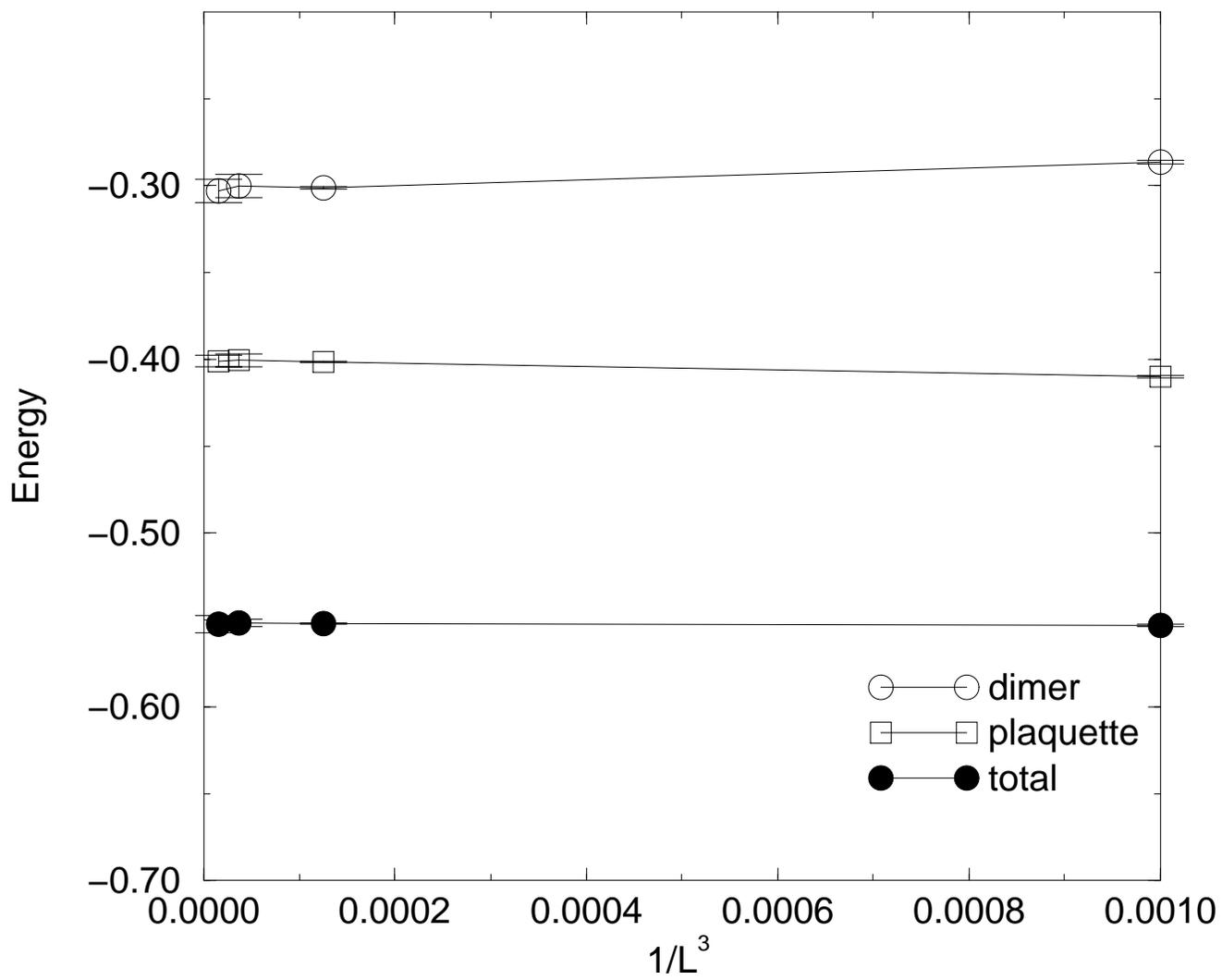
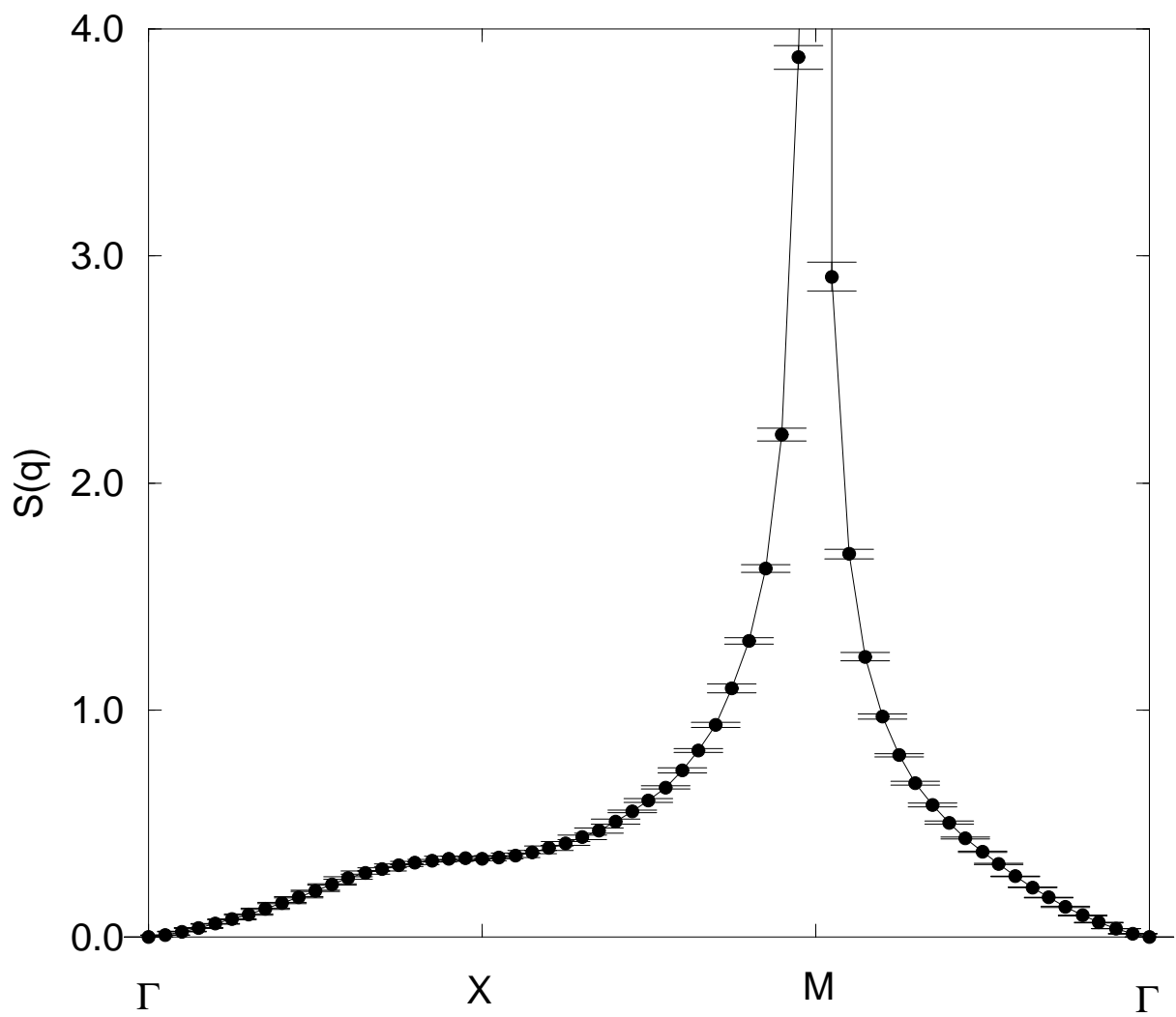
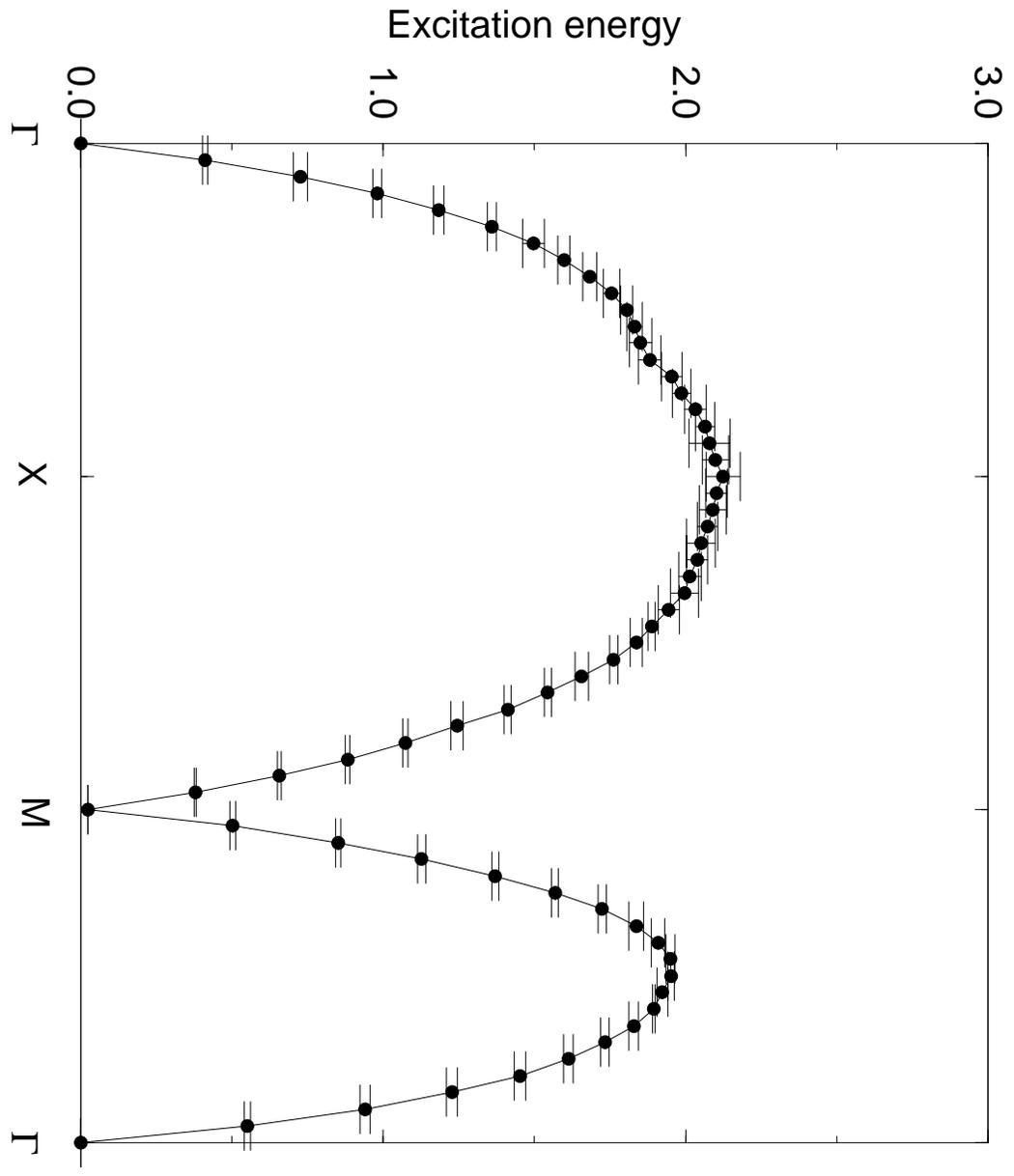
\end{document}